\documentclass[journal, onecolumn]{IEEEtran}
\usepackage{ifpdf}

\usepackage{cite}

\ifCLASSINFOpdf
  \usepackage[pdftex]{graphicx}
\else
\fi
\usepackage{amsmath}
\usepackage{bm}
\usepackage{amssymb}
\usepackage{mathtools}

\usepackage{url}

\newtheorem{theo}{Theorem}
\newtheorem{prop}{Proposition}
\newtheorem{defi}{Definition}
\newtheorem{lemm}{Lemma}

\newtheorem{cor}{Corollary}
\newtheorem{assum}{Assumption}

\begin{document}
%
\title{A Stochastic Model for Block Segmentation of Images Based on the Quadtree and the Bayes Code for It}
%
%
%

\author{Yuta~Nakahara and
        Toshiyasu~Matsushima,~\IEEEmembership{Member,~IEEE,}
\thanks{Y. Nakahara is with the Center for Data Science, Waseda University, 1-6-1 Nishiwaseda, Shinjuku-ku, Tokyo, 162-8050, Japan.}
\thanks{T. Matsushima is with the Department of Pure and Applied Mathematics, Waseda University, 3-4-1 Okubo, Shinjuku-ku, Tokyo, 169-8555, Japan.}
\thanks{Manuscript received April 19, 2005; revised August 26, 2015. This paper was presented in part at the 2020 Data Compression Conference (DCC2020)\cite{DCC}.}}

%
%

\markboth{Journal of \LaTeX\ Class Files,~Vol.~14, No.~8, August~2015}%
{Shell \MakeLowercase{\textit{et al.}}: Bare Demo of IEEEtran.cls for IEEE Journals}
%



\maketitle

\begin{abstract}
In information theory, lossless compression of general data is based on an explicit assumption of a stochastic generative model on target data. However, in lossless image compression, the researchers have mainly focused on the coding procedure that outputs the coded sequence from the input image, and the assumption of the stochastic generative model is implicit. In these studies, there is a difficulty in confirming the information-theoretical optimality of the coding procedure to the stochastic generative model. Hence, in this paper, we propose a novel stochastic generative model of images by redefining the implicit stochastic generative model in a previous coding procedure. That is based on the quadtree so that our model effectively represents the variable block size segmentation of images. Then, we construct the Bayes code optimal for the proposed stochastic generative model. In general, the computational cost to calculate the posterior distribution required in the Bayes code increases exponentially for the image size. However, we introduce an efficient algorithm to calculate it in the polynomial order of the image size without loss of the optimality. Some experiments are performed to confirm the flexibility of the proposed stochastic model and the efficiency of the introduced algorithm.
\end{abstract}


%
\IEEEpeerreviewmaketitle

\section{Introduction}
%
%
%
%

\subsection{Lossless data compression in information-theory}

\IEEEPARstart{I}{n} information theory, lossless compression for general data (not only images) is based on an explicit assumption of a \emph{stochastic generative model} $p(\bm x)$ on target data $\bm x$\cite{Shannon}. This assumption determines the theoretical limit, which is called entropy, of the expected code length for $p(\bm x)$. When $p(\bm x)$ is known, the entropy codes like Huffman code\cite{Huffman} and arithmetic code (see, e.g., \cite{arithmetic}) achieve the theoretical limit. Then, the researchers have considered a set-up in which $p(\bm x)$ is unknown. One method to describe the uncertainty of $p(\bm x)$ is considering a class of parameterized stochastic generative models $p(\bm x | \bm \theta)$ and assuming the class is known but the parameter $\bm \theta$ is unknown. Even for this set-up, the researchers have proposed a variety of stochastic generative model classes and coding algorithms achieving those theoretical limits, e.g., i.i.d. model class, Markov model class, context tree model class, and so on (see, e.g., \cite{Davisson, enumerative, CT, CTW, kontoyiannis}). 

In this set-up, the variety of the stochastic generative model is described as that of unknown parameters or model variables. For example, the i.i.d. model can be determined by a vector $\bm \theta$ whose elements are occurrence probabilities of each symbol and described as $p(\bm x | \bm \theta)$. Markov model contains another variable $s$ that represents the state or context, which is a string of most recent symbols at each time point, and the occurrence probability vector $\bm \theta_s$ is multiplied for each $s$. Then, the Markov model can be described as $p(\bm x| \bm \theta_s, s)$. Further, when the order of Markov model is unknown, that contains another variable $k$ which represents the order and the occurrence probability $\bm \theta_s^k$ and the state variable $s^k$ are multiplied for each $k$. Then, the Markov model with unknown order can be described as $p(\bm x | \bm \theta_s^k, s^k, k)$. Moreover, in the context tree model, the order depends on context and $k$ is replaced by an unknown model variable $m$ that represents a set of contexts. Finally, the context tree model can be described as $p(\bm x | \bm \theta_s^m, s^m, m)$.

It should be noted that these parameters and model variable $\bm \theta$, $k$, $m$ are the \emph{statistical parameters} that govern the generation of the data $\bm x$. Therefore, the optimal coding algorithm for these stochastic generative models inevitably contains the optimal estimation $\hat{\bm \theta}(\bm x)$, $\hat{k}(\bm x)$, $\hat{m}(\bm x)$ of them as a sub-routine\footnote{In a Bayesian setting, they can be estimated as posteriors $p(\bm \theta | \bm x)$, $p(k | \bm x)$ or $p(m | \bm x)$}. The explicit assumption of the stochastic generative model and the construction of the coding algorithm with the optimal parameter estimation have been successful in the text compression. In fact, various text coding algorithms have been derived (e.g., \cite{CT, CTW, kontoyiannis}).

\subsection{Lossless image compression as a image processing}

However, in most cases of lossless ``image'' compression, the main focus is on the construction of the \emph{coding procedure} $f(\bm x)$ that just outputs the coded sequence from the input pixel values $\bm x$ without explicit assumption of a stochastic generative model. In the usual case, the coding algorithm has a \emph{tuning parameter} $a$ and represented as $f(\bm x; a)$. This tuning parameter $a$ is tuned adaptive to pixel values $\bm x$ and we express this tuning method as $\tilde{a}(\bm x)$. Then, the coded sequence $f(\bm x; \tilde{a}(\bm x))$ from $\bm x$ is uniquely determined. 

Therefore, the variety of the coding procedures is described as that of the tuning parameters and the tuning methods. More specifically, we give a brief review of a type of lossless image coding called predictive coding. Most of the predictive coding procedure have the form $f(x^{t-1}; a, b)$ with two parameters $a$ and $b$. $a$ is a parameter of the predictor, which predicts the next pixel value $x_t$ from the already compressed pixels $x^{t-1}$ at time $t$. $b$ is a parameter of the coding probability (not a occurrence probability in general), which is assigned to the predictive error sequence. Then, the predictive error sequence and the coding probability are input to the entropy codes like the arithmetic code\cite{arithmetic}. For example, in JPEG-LS\cite{JPEGLS}, they use three predictors that are switched according to the neighboring pixels. This can be regarded that $a \in \{ 1, 2, 3 \}$ corresponds to the index of the three predictors and the rule to switch them is represented by $\tilde{a}(x^{t-1})$. The coding probability of JPEG-LS\cite{JPEGLS} is represented by a two-sided geometric distribution, which is tuned by the past sequence $x^{t-1}$. This can be regarded that $b$ is a parameter of the two-sided geometric distribution and $\tilde{b}(x^{t-1})$ is the tuning method of it. In other studies \cite{kuroki, 2DAR, Glicbawls, WLS, BayesLR, Vanilc}, they have proposed coding procedures $f(x^{t-1}; a, b, \bm c_a)$ in which coefficients $\bm c_a$ of each linear predictor are tuned by a certain method $\tilde{\bm c}_a(x^{t-1})$, e.g., least squares method or weighted least squares method. In \cite{TMW, Bayes_avg}, they proposed coding procedures $f(x^{t-1}; a, b, \bm c_a, \bm w)$ in which multiple predictors are combined according to another tuning parameter $\bm w$ represents the weights of each predictor. Regarding the coding probability, the study \cite{MRP} deals with a procedure $f(x^{t-1}; a, b, \bm c_a, d)$ in which coding probability is represented by the generalized Gauss distribution that has another tuning parameter $d$.\footnote{These notation is just for the explanation of the idea of the previous studies; It does not completely matches the notation of each paper, and it does not contain all of the tuning parameters of each procedure.} One of the latest study constructing a complicate coding procedure is \cite{ulacha}, in which numerous tuning parameters are tuned through the careful experiments. Lossless image compression using deep learning (see, e.g., \cite{deep_learning}) can be regarded as one of the coding procedure with a huge number of tuning parameters that are pre-trained.

These studies have been practically successful. However, it should be noted that the tuning  parameters $a$ and $b$ are not the statistical parameters that govern the generation of pixel values $\bm x$ since they are introduced just to add a degree of freedom to the coding procedure. Even the parameter $b$, which superficially looks a parameter of a probability distribution, does not directly govern the generation of pixel values $\bm x$ unless the coding procedure is extremely simple; it is just used to represent the coding probability with less number of variables. Therefore, the tuning of these parameters adaptive to $\bm x$ is not theoretically grounded by the statistics nor information theory. If our task was not the lossless compression, e.g., lossy compression, image hyperresolution, and so on, this parameter tuning would be evaluated from various point of view, e.g., subjective evaluation by human. It is because such tasks have difficulty in performance measure itself. Besides, in lossless image compression, it should be evaluated from information-theoretical perspective. These parameters should be tuned to decrease the difference between the expected code length and the entropy of assumed stochastic generative model, and we have to say any other tuning methods are heuristic unless they pursue the added value except for the coding rate. However, such an information-theoretical evaluation is impossible because there is no explicit assumption of the stochastic generative model $p(\bm x)$ and the entropy ---theoretical limit of the expected code length--- itself is not defined. This is a critical problem of the previous studies above. In addition, the more the tuning parameters are introduced, the more difficult the construction of the tuning method becomes, since there is no confirmation of the optimality of each tuning method. 

\subsection{Lossless image compression on an explicitly redefined the stochastic generative model}

However, there are some coding procedures $f(\bm x; a)$\cite{JPEGLS, CALIC, kuroki, 2DAR, BayesLR, Glicbawls, Bayes_avg, Vanilc, TMW, MRP} whose tuning parameter $a$ can be regarded as a statistical parameter of an implicitly assumed statistical generative model $p(\bm x | a)$ by changing the viewpoint\footnote{In some of these studies, the assumption of the stochastic generative model are claimed, but the distinguishment of the stochastic generative model from the coding probability is ambiguous, and the discussion of the information-theoretical optimality is insufficient.}. Further, its parameter tuning method $\tilde{a}(\bm x)$ could be regraded as an heuristic approximation of the information-theoretically optimal estimation $\hat{a}(\bm x) \approx \tilde{a}(\bm x)$. Then, explicitly redefining the implicit stochastic generative model behind the previous coding procedures, we can construct a statistical generative model supported by the practical achievements. Moreover, if we derive the information-theoretically optimal coding algorithm for the extended stochastic generative model, this algorithm inevitably contains the optimal parameter estimation $\hat{a}(\bm x)$, which is the improved version of $\tilde{a}(\bm x)$ with information-theoretical optimality.

To derive such a coding algorithm, we can utilize the coding algorithms in the text coding. Although image data is different from the the text data, their stochastic generative models may contain a similar structure and we may utilize the parameter estimation method in the text coding. In fact, we utilize the efficient algorithm for the context tree model class\cite{CT, CTW, kontoyiannis} for our stochastic generative model in this paper.

It is true that the coding algorithm constructed in this approach does not necessarily work for real images, since the optimality is guaranteed only for the stochastic generative model and it is difficult to prove that the real images generated from the assumed stochastic generative model. Therefore, the constructed coding algorithm might be inferior to the existing one in the initial stage of this approach. However, our claim is that this problem should not be solved by a heuristic tuning of parameter in the coding procedure but an explicit extension of the stochastic generative model, as mach as possible. Such a parameter tuning should be done in the final stage before the implementation or standardization.

The previous studies based on this approach are \cite{ISITA} and \cite{IWAIT}. In \cite{ISITA}, they proposed a two-dimensional autoregressive model and the optimal coding algorithm by interpreting the basic procedure\cite{JPEGLS, CALIC, kuroki, 2DAR, BayesLR} of the predictive coding as a stochastic generative model. In \cite{IWAIT}, they proposed a two-dimensional autoregressive hidden Markov model by interpreting the predictor weighting procedure around a diagonal edge\cite{TMW} as a stochastic generative model. However, these stochastic generative models do not have enough flexibility to represent the non-stationarity among segments of an image.

\subsection{The contribution of this paper}

Then, our target data are the images in which the properties of pixel values are different depending on the segments. In this paper, we achieve the following purposes.
\begin{enumerate}
\item To propose a stochastic generative model that effectively represents the non-stationarity among the segments in an image.
\item To derive an information-theoretically optimal code for the proposed stochastic model.
\item To derive an efficient algorithm to implement the code without loss of the optimality.
\end{enumerate}

A trivial way to represent the non-stationarity as a stochastic generative model is to divide the image into fixed-size blocks and to assume different probability distributions for each block. However, such a stochastic generative model is not flexible enough to represent the smaller segments and inefficient to represent the larger segments than the block size.

On the other hand, one of the most efficient lossless image coding procedure of \cite{MRP} contains a preprocessing to determine a quadtree that represents a variable block size segmentation. Then, different predictors are assigned to each block to mitigate the non-stationality. The quadtree is also used in various fields of image and video processing to represent the variable block size segmentation, and its flexibility and computational efficiency  are reported by a number of studies, e.g., in H.265 \cite{H265}. However, the quadtree in these studies is a tuning parameter of a procedure. There are no studies that regard the quadtree as a statistical model variable $m$ of a stochastic generative model $p(\bm x | m)$ which govern the generation of pixel values $\bm x$ and construct the information-theoretically optimal code for it, to the best of our knowledge.

In this paper, we propose a novel stochastic generative model based on the quadtree, so that our model effectively represents the non-stationarity among segments by the variable block size segmentation. Then, we construct the information-theoretically optimal code for the proposed stochastic generative model. Moreover, we introduce a computationally efficient algorithm to implement our code without loss of optimality, taking in the knowledge of the text coding\cite{CT, CTW, kontoyiannis}.

Although the main theme of this paper is lossless image compression, the substantial contribution of our results is the construction of the stochastic model. Therefore, the proposed stochastic model contributes to not only lossless image compression but also any other stochastic image processing like recognition, generation, feature extraction, and so on.

The organization of this paper is as follows. In Section 2, we describe the proposed stochastic generative model. In Section 3, we derive the optimal code for the proposed model. In Section 4, we derive an efficient algorithm to implement the derived code. In Section 5, we perform some experiments to confirm the flexibility of our stochastic generative model and the efficiency of our algorithm. In Section 6, we describe future works. Section 7 is the conclusion of this paper.

\section{The proposed stochastic model}
Let $\mathcal{V}$ denote a set of possible values of a pixel. For example, $\mathcal{V} = \{ 0, 1 \}$ for binary images, $\mathcal{V} = \{ 0, 1, \dots , 255 \}$ for gray scale images, and $\mathcal{V} = \{ 0, 1, \dots , 255 \}^3$ for color images. Let $\mathbb{N}$ denote the set of natural numbers. Let $h \in \mathbb{N}$ and $w \in \mathbb{N}$ denote a height and a width of a image, respectively. Although our model is able to represent any rectangular images and its block segmentation, we assume that $h = w = 2^{d_\mathrm{max}}$ for $d_\mathrm{max} \in \mathbb{N}$ in the following for the simplicity of the notation. Then, let $V_t$ denote the random variable of the $t$-th pixel value in order of the raster scan, and $v_t \in \mathcal{V}$ denote its realized value. 
Note that $V_t$ is at $x(t)$-th row and $y(t)$-th column, where $t = x(t) w + y(t)$ and $y(t) < w$. In addition, let $V^t$ denote the sequence of pixel values $V_0, V_1, \dots , V_t$. Note that all the indices start from zero in  this paper.


We consider the pixel value $V_t$ is generated from various probability distributions depending on a model $m \in \mathcal{M}$ and parameters $\bm \theta^m \in \bm \Theta^m$. Therefore, they are represented by $p(v_t | v^{t-1}, \bm \theta^m, m)$ in general.
Note that the model $m$ and the parameters $\bm \theta^m$ are unobservable and should be estimated in actual situations. The definitions of $m$ and $\bm \theta^m$ are as follows.

\begin{defi}
Let $s_{(x_1 y_1) (x_2 y_2) \cdots  (x_d y_d)}$ denote the following index set called ``block''
\begin{align}
&s_{(x_1 y_1) (x_2 y_2) \cdots (x_d y_d)} \coloneqq \left\{ (i, j) \in \mathbb{Z}^2 \; \middle| \; \sum_{d' = 1}^d \frac{x_{d'}}{2^{d'}} \leq \frac{i}{2^{d_\mathrm{max}}} < \left( \sum_{d' = 1}^d \frac{x_{d'}}{2^{d'}} + \frac{1}{2^d} \right), \sum_{d' = 1}^d \frac{y_{d'}}{2^{d'}} \leq \frac{j}{2^{d_\mathrm{max}}} < \left( \sum_{d' = 1}^d \frac{y_{d'}}{2^{d'}} + \frac{1}{2^d} \right) \right\} ,
\end{align}
where $x_{d'}, y_{d'} \in \{ 0, 1\}$, $d \leq d_\mathrm{max}$, and $\mathbb{Z}$ denotes the set of integers. In addition, let $s_{\lambda}$ be the set of whole indices $s_{\lambda} \coloneqq \{ 0, 1, \dots h-1 \} \times \{ 0, 1, \dots , w-1 \}$. Then, let $\mathcal{S}$ denote the set which consists of all the above index sets, namely $\mathcal{S} \coloneqq \{ s_\lambda, s_{(0 0)}, \dots , s_{(1 1)}, s_{(0 0) (0 0)}, \dots , s_{(1 1) (1 1)}, \dots , s_{(1 1) (1 1) \cdots (1 1)}  \}$.
\end{defi}

\begin{figure*}[t]
\centering
\includegraphics[width=5in, clip]{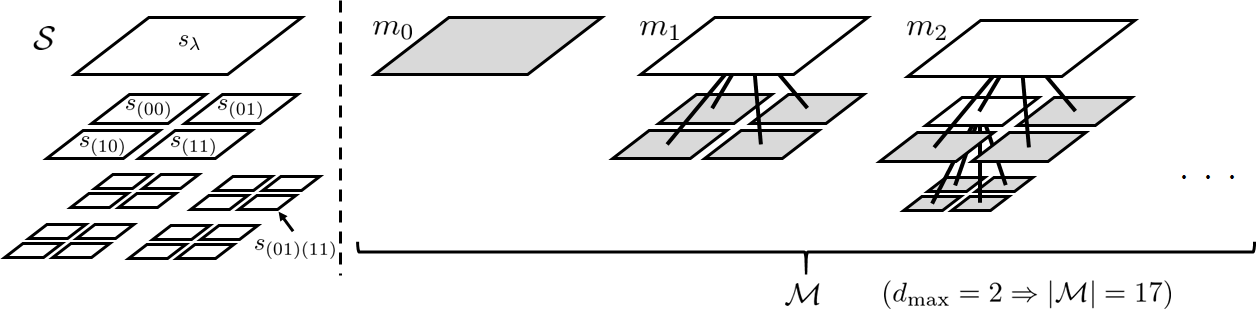}
\caption{An example of node set $\mathcal{S}$ and models $m$.} 
\label{model} 
\end{figure*}

\begin{defi}
We define the model $m$ as a full quadtree whose nodes are elements of $\mathcal{S}$. Let $\mathcal{L}^m \subset \mathcal{S}$ and $\mathcal{I}^m \subset \mathcal{S}$ denote the set of the leaf nodes and the inner nodes of $m$, respectively. Then, $\mathcal{L}^m$ corresponds to a pattern of variable block size segmentation, as shown in Fig.\ \ref{model}. Let $\mathcal{M}$ denote the set of full quadtrees whose depth is smaller than or equal to $d_\mathrm{max}$.
\end{defi}


\begin{defi}
Each leaf node $s \in \mathcal{L}^m$ of the model $m$ has a parameter $\theta_s^m$ whose parameter space is $\Theta_s^m$. We define $\bm \theta^m$ as a tuple of parameters $\{ \theta_s^m \}_{s \in \mathcal{L}^m}$, and let $\bm \Theta^m$ denote the total parameter space of them. 
\end{defi}

Under the model $m \in \mathcal{M}$ and the parameters $\bm \theta^m \in \bm \Theta^m$, we assume that the $t$-th pixel value $v_t \in \mathcal{V}$ is generated as follows.
\begin{assum}\label{out_prob}
We assume that
\begin{align}
p(v_t | v^{t-1}, \bm \theta^m, m) = p(v_t | v^{t-1}, \theta_{s}^m), \label{generative_model}
\end{align}
where $s \in \mathcal{L}^m$ satisfies $(x(t), y(t)) \in s$.
\end{assum}
Thus, the pixel value $V_t$ depends only on the parameter of the block $s$ which contains $V_t$ under the past sequence $V^{t-1}$.

\section{The Bayes code for the proposed model}
If we know the true model $m$ and the parameters $\bm \theta^m$, we are able to compress the pixel value $v_t$ up to the entropy of $p(v_t | v^{t-1}, \bm \theta^m, m)$ by the well-known entropy code like the arithmetic code. However, the true $m$ and $\bm \theta^m$ are unobservable. One reasonable solution is to estimate them and substitute the estimated ones $\hat{m}$ and $\hat{\bm \theta}^m$ into $p(v_t | v^{t-1}, \bm \theta^m, m)$. Then, we can use $p(v_t | v^{t-1}, \hat{\bm \theta}^m, \hat{m})$ as a coding probability of the entropy code.

However, there is another powerful solution, in which we assume prior distributions $p(m)$ and $p(\bm \theta^m | m)$. Then, we estimate the true coding probability $p(v_t | v^{t-1}, \bm \theta^m, m)$ itself instead of $m$ and $\bm\theta^m$ by $q (v_t | v^{t-1})$ so that $q (v_t | v^{t-1})$ can minimize the \textit{Bayes risk function} based on the \textit{loss function} between the expected code length of entropy code using $p(v_t | v^{t-1}, \bm \theta^m, m)$ and that using $q (v_t | v^{t-1})$. The code constructed by such a method is called the Bayes code (see, e.g., \cite{Bayes} and \cite{Bayes_Analysis}).

It is known that the expected code length of the Bayes code converges to the entropy of the true stochastic model for sufficiently large data length $t$, and its convergence speed achieves the theoretical limits\cite{Bayes_Analysis}. In fact, the Bayes code achieves remarkable performances in text compression (e.g., \cite{CT}).

Therefore, we derive the Bayes code for the proposed stochastic model. According to the general formula in \cite{Bayes}, the optimal coding probability for $v_t$ in the scheme of the Bayes code is derived as follows:
\begin{prop}\label{ideal}
The optimal coding probability $q^* (v_t | v^{t-1})$ which minimizes the Bayes risk function is
\begin{align}
&q^* (v_t | v^{t-1}) =  p(v_t | v^{t-1}) = \sum_{m \in \mathcal{M}} p(m | v^{t-1}) \int p(v_t | v^{t-1}, \bm \theta^m, m) p(\bm \theta^m | v^{t-1}, m) \mathrm{d}\bm \theta^m. \label{BC}
\end{align}
We call $q^* (v_t | v^{t-1})$ the Bayes optimal coding probability.
\end{prop}

Proposition \ref{ideal} implies that we should calculate the posterior distributions $p(m | v^{t-1})$ and $p(\bm \theta^m | v^{t-1}, m)$. Then, we should use the coding probability which is a weighted mixture of $p(v_t | v^{t-1}, \bm \theta^m, m)$ for every block segmentation pattern $m$ and parameters $\bm \theta^m$ according to the posteriors $p(m | v^{t-1})$ and $p(\bm \theta^m | v^{t-1}, m)$.

\section{The efficient algorithm to calculate the coding probability}\label{sec_alg}
Unfortunately, the Bayes optimal coding probability (\ref{BC}) contains computationally difficult calculations. As the depth $d_\mathrm{max}$ of full quadtree increases, the amount of calculation for the sum with respect to $m \in \mathcal{M}$ increases exponentially. Moreover, the posterior $p(m | v^{t-1})$ does not have a closed-form expression in general.\footnote{Strictly speaking, a few problems are also left. Both of the integral with respect to $\bm \theta^m$ and the posterior $p(\bm \theta^m | m, v^{t-1})$ do not have closed-form expressions in general. These problems can be solved in various methods depending on the setting of $p(v_t | v^{t-1}, \bm \theta^m, m)$ and $p(\bm \theta^m | m)$ and almost independent of our proposed model. Therefore, we just describe an example of a feasible setting of $p(v_t | v^{t-1}, \bm \theta^m, m)$ and $p(\bm \theta^m | m)$ in the next section.}

Similar problems are studied in the text compression and efficient algorithms to calculate the coding probability is constructed (see, e.g., \cite{CT, CTW, kontoyiannis}). In these algorithms, the weighted sum of the context trees is calculated instead of the quadtrees. We apply it for our proposed model. In this section, we focus to describe the procedure of the constructed algorithm. Its validity is described in Appendix \ref{validity}.

First, we assume the following priors on $m$ and $\bm \theta^m$.
\begin{assum}\label{m_prior}
We assume that each node $s \in \mathcal{S}$ has a hyperparameter $g_s \in [0, 1]$, and the model prior $p(m)$ is represented by
\begin{align}
p(m) = \prod_{s \in \mathcal{L}^m} (1-g_s) \prod_{s' \in \mathcal{I}^m} g_{s'}, \label{pm}
\end{align} 
where $g_s = 0$ for $s$ whose cardinality $|s|$ equals to 1.
\end{assum}
The idea of this form is to represent $p(m)$ as a product of the probability that the block $s$ is divided. Such a probability is denoted by $g_s$ in (\ref{pm}).
A proof that the above prior satisfies the condition $\sum_{m \in \mathcal{M}} p(m) = 1$ is in Appendix \ref{validity}. Note that the above assumption does not restrict the expressive capability of the general prior in the meaning that each model $m$ still has possibility to be assigned a non-zero probability $p(m) > 0$.

\begin{assum}\label{param}
For each model $m \in \mathcal{M}$, we assume that
\begin{align}
p(\bm \theta^m | m) = \prod_{s \in \mathcal{L}^m} p(\theta_s^m | m).
\end{align}
Moreover, for any $m, m' \in \mathcal{M}$, $s \in \mathcal{L}^m \cap \mathcal{L}^{m'}$, and $\theta_s \in \Theta_s$, we assume that
\begin{align}
p(\theta_s | m) = p(\theta_s | m') \eqqcolon p_s (\theta_s).
\end{align}
\end{assum}
Therefore, each element $\theta_s^m$ of the parameters $\bm \theta^m$ depends only on $s$ and independent both of the other elements and the model $m$.

From Assumptions \ref{out_prob} and \ref{param}, the following lemma holds.
\begin{lemm}\label{param_posterior}
For any $m, m' \in \mathcal{M}$, $s \in \mathcal{L}^m \cap \mathcal{L}^{m'}$, and $v^t \in \mathcal{V}^t$, if $(x(t), y(t)) \in s$, then
\begin{align}
p(v_t | v^{t-1}, m) 
&= p(v_t | v^{t-1}, m').
\end{align}
Then, we represent it by $\tilde{q}(v_t | v^{t-1}, s)$ because it depends on not $m$ but $s$. 
\end{lemm}

The proof of Lemma \ref{param_posterior} is in Appendix \ref{validity}. Lemma \ref{param_posterior} means that the optimal coding probability for $v_t$ depends only on the block $s$ which contains $v_t$, and it could be calculated as $q(v_t | v^{t-1}, s)$ if $s$ was known.

At last, the efficient algorithm to compute the Bayes optimal coding probability $q^* (v_t | v^{t-1})$ is represented as an iteration of updating $g_s$ and summing the functions $\tilde{q}(v_t | v^{t-1}, s)$ weighted by $g_s$ for nodes on a path of the complete quadtree on $\mathcal{S}$. 

\begin{figure*}[t]
\centering
\includegraphics[width=4in, clip]{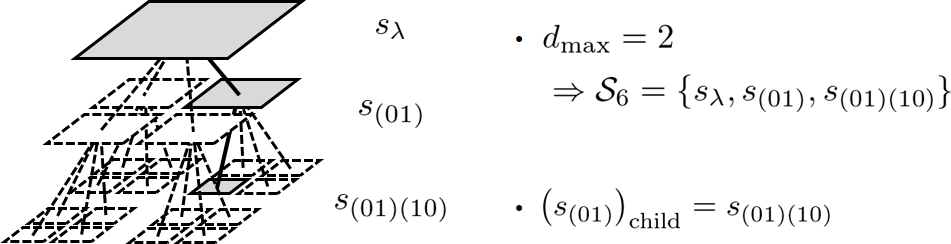}
\caption{An example of a path constructed from $\mathcal{S}_t$.} 
\label{path} 
\end{figure*}

\begin{defi}
Let $\mathcal{S}_t$ denote the set of nodes which contain $(x(t), y(t))$. They construct a path from the leaf node \\ $s_{(x_1y_1) (x_2 y_2) \cdots (x_{d_\mathrm{max}} y_{d_\mathrm{max}})} = \{ (x(t), y(t)) \}$ to the root node $s_\lambda$ on the complete quadtree whose depth is $d_\mathrm{max}$ on $\mathcal{S}$, as shown in Fig.\ \ref{path}.
In addition, let $s_\mathrm{child} \in \mathcal{S}_t$ denote the child node of $s \in \mathcal{S}_t$ on that path.
\end{defi}

\begin{defi}
We define the following recursive function $q(v_t | v^{t-1}, s)$ for $s \in \mathcal{S}_t$.
\begin{align}
&q(v_t | v^{t-1}, s) \coloneqq
\begin{cases}
\tilde{q}(v_t | v^{t-1}, s), & |s| = 1,\\
(1-g_{s | t-1}) \tilde{q}(v_t | v^{t-1}, s) + g_{s | t-1} q(v_t | v^{t-1}, s_{\mathrm{child}}), & \mathrm{otherwise},
\end{cases}\label{recursion}
\end{align}
where $g_{s | t}$ is also recursively updated as follows. 
\begin{align}
&g_{s | t} \coloneqq
\begin{cases}
g_s, & t = -1\\
g_{s|t-1}, & t  \geq  0 \wedge (s  \notin  \mathcal{S}_t \vee |s| = 1)\\
\frac{g_{s | t-1} q(v_t | v^{t-1}, s_{\mathrm{child}})}{q(v_t | v^{t-1}, s)}, & t  \geq  0 \wedge s  \in  \mathcal{S}_t \wedge |s|  >  1.
\end{cases}\label{update}
\end{align}
\end{defi}

Then, the following theorem holds.
\begin{theo}\label{alg}
The Bayes optimal coding probability $q^* (v_t | v^{t-1})$ for the proposed model is calculated by
\begin{align}
q^* (v_t | v^{t-1}) = q(v_t | v^{t-1}, s_\lambda).
\end{align}
\end{theo}
The proof of Theorem \ref{alg} is in Appendix \ref{validity}. Theorem \ref{alg} means that the summation with respect to $m \in \mathcal{M}$ in (\ref{BC}) is able to be replaced by the summation with respect to $s \in \mathcal{S}_t$ and it costs only $O(d_\mathrm{max})$. In a sense, $(1-g_{s|t-1})$ can be regarded as the marginal posterior probability that the true block division was stopped at $s$. Then the proposed algorithm takes a mixture of the coding probability $\tilde{q}(v_t | v^{t-1}, s)$, weighting such a case with $(1-g_{s|t-1})$ and the other cases with $g_{s|t-1}$.

\section{Experiments}
In this section, we perform two experiments. The purpose of the first experiment is to confirm the Bayes optimality of $q(v_t | v^{t-1}, s_\lambda)$. Therefore, we use synthetic images randomly generated from the proposed model. The purpose of the second experiment is to demonstrate the flexibility of our model. Therefore, we use a well-known benchmark image. We also use the Bayes optimal code for fixed block size segmentation\footnote{Let $2^d$ be the fixed block size. Such a model is derived by substituting $g_s = 1$ for $s$ whose depth is smaller than $d_\mathrm{max} - d$ and $g_s = 0$ otherwise.} for comparison in both experiments.

In the following, we assume $\mathcal{V} = \{ 0, 1 \}$. In other words, we treat only binary images. $p(v_t | v^{t-1}, \bm \theta^m, m)$ is assumed to be Bernoulli distribution $\mathrm{Bern} (v_t | \theta^m_s)$ for $s$ which satisfies $(x(t), y(t)) \in s$. Each element of $\bm \theta^m$ is i.i.d.\ distributed with Beta distribution $\mathrm{Beta} (\theta | \alpha, \beta)$, which is the conjugate distribution of Bernoulli distribution. Therefore, the integral in (\ref{BC}) has a closed-form. The hyperparameter $g_s$ of the model prior is $g_s = 1/2$ for every $s \in \mathcal{S}$, and the hyperparameters of the Beta distribution are $\alpha = \beta = 1/2$.

\subsection{Experiment 1}
The setting of Experiment 1 is as follows. The width and height of images are $w = h =$ $2^{d_\mathrm{max}} = 64$. Then, we generate 1000 images according to the following procedure.
\begin{enumerate}
\item Generate $m$ according to (\ref{pm}).
\item Generate $\theta^m_s$ according to $p(\theta_s^m | m)$ for $s \in \mathcal{L}^m$.
\item Generate pixel value $v_t$ according to $p(v_t | v^{t-1}, \bm \theta^m, m)$ for $t \in \{ 0, 1, \dots , hw-1 \}$.
\item Repeat 1) to 3) in 1000 times.
\end{enumerate}
Then, we compress these 1000 images. The size of the image is saved in the header of the compressed file using 4 bytes. The coding probability calculated by the proposed algorithm is quantized in $2^{16}$ levels and substituted into the range coder \cite{range_coder}.

The coding rates (bit/pel) averaged over all the images are shown in Table \ref{tb_bayes}. Our proposed code has the minimum coding rate as expected by the Beyes optimality.

\begin{table}[t]
\begin{center}
\caption{\label{tb_bayes}%
The average coding rates (bit/pel)}
{
\renewcommand{\baselinestretch}{1}\footnotesize
\begin{tabular}{|c|c|c|c|}
\hline
Quadtree (proposed) & Fixed size 4 & Fixed size 8 & Fixed size 16\\
\hline
\textbf{0.619} & 0.705 & 0.659 & 0.679\\
\hline
\end{tabular}}
\end{center}
\end{table}

\subsection{Experiment 2}
\begin{figure*}[t]
  \begin{center}
    \begin{tabular}{c}

      \begin{minipage}{0.5\hsize}
        \begin{center}
          \includegraphics[width=2in,clip]{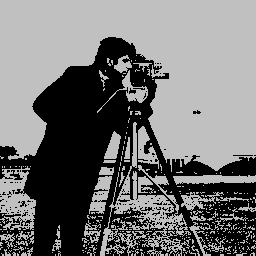}
        \end{center}
      \end{minipage}

      \begin{minipage}{0.5\hsize}
        \begin{center}
          \includegraphics[width=2in,clip]{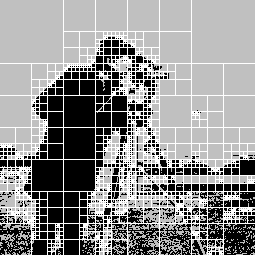}
        \end{center}
      \end{minipage}

    \end{tabular}
    \caption{The original image (left) and the MAP estimated model $m^{\mathrm{MAP}}$.}
    \label{posterior}
  \end{center}
\end{figure*}

In Experiment 2, we compress the binarized version of \url{camera.tif} from \cite{Waterloo}, where the threshold of binarization is 128. The setting of the header and the range coder is the same as those of Experiment 1. Figure \ref{posterior} visualizes the maximum a posteriori (MAP) estimation $m^{\mathrm{MAP}} = \mathrm{arg} \max_{m} p(m | v^{hw-1})$, which is calculated as a by-product of the compression by the algorithm detailed in Appendix \ref{max_algorithm}. It shows that our proposed model has the flexibility to represent the non-stationarity among the regions. The coding rate for \url{camera.tif} is shown in Table \ref{tb_camera}, and it implies that our code has a certain performance for real images.

\begin{table}[t]
\begin{center}
\caption{\label{tb_camera}%
The coding rates for the \protect\url{camera.tif} from \cite{Waterloo}(bit/pel)}
{
\renewcommand{\baselinestretch}{1}\footnotesize
\begin{tabular}{|c|c|c|c|}
\hline
Quadtree (proposed) & Fixed size 4 & Fixed size 8 & Fixed size 16\\
\hline
\textbf{0.323} & 0.427 & 0.388 & 0.430\\
\hline
\end{tabular}}
\end{center}
\end{table}

\section{Future works}
In this paper, we focused only on the stochastic representation of the non-stationarity among the segments. The discussion about the stochastic model $p(v_t | v^{t-1}, \bm \theta^m, m)$ and the prior $p(\bm \theta^m | m)$ to be assumed in each block is out of the scope. This is the first future work. 
For example, our model also works on the pairs of categorical distribution and Dirichlet distribution, normal distribution and normal-gamma distribution, and 2-dimensional autoregressive model and normal-gamma distribution\cite{ISITA}. Moreover, using an approximative Bayesian estimation like the variational Bayesian method, we expect that more complicated stochastic models like \cite{IWAIT} are able to be assumed.

The second future work is to apply our model to other stochastic image processing, namely, image recognition, image generation, image inpainting, future extraction, and so on. In particular, image generation and image inpainting may be suitable because the whole structure of stochastic image generation is described in our model and the parameters of the stochastic model are able to be learned optimally.

\section{Conclusion}
We proposed a novel stochastic model based on the quadtree, so that our model effectively represents the variable block size segmentation of images. Then, we constructed a Bayes code for the proposed stochastic model. Moreover, we introduced an efficient algorithm to implement it in polynomial order of data size without loss of optimality. Some experiments both on synthetic and real images demonstrated the flexibility of our stochastic model and the efficiency of our algorithm.

\section*{Acknowledgement}
We would like to thank the members of Matsushima laboratory for meaningful discussions.
This work was supported by JSPS KAKENHI Grant Numbers JP17K06446, JP18K11585, and JP19K04914.

\bibliographystyle{IEEEtran}
\bibliography{refs.bib}

\begin{thebibliography}{10}
\providecommand{\url}[1]{#1}
\csname url@samestyle\endcsname
\providecommand{\newblock}{\relax}
\providecommand{\bibinfo}[2]{#2}
\providecommand{\BIBentrySTDinterwordspacing}{\spaceskip=0pt\relax}
\providecommand{\BIBentryALTinterwordstretchfactor}{4}
\providecommand{\BIBentryALTinterwordspacing}{\spaceskip=\fontdimen2\font plus
\BIBentryALTinterwordstretchfactor\fontdimen3\font minus
  \fontdimen4\font\relax}
\providecommand{\BIBforeignlanguage}[2]{{%
\expandafter\ifx\csname l@#1\endcsname\relax
\typeout{** WARNING: IEEEtran.bst: No hyphenation pattern has been}%
\typeout{** loaded for the language `#1'. Using the pattern for}%
\typeout{** the default language instead.}%
\else
\language=\csname l@#1\endcsname
\fi
#2}}
\providecommand{\BIBdecl}{\relax}
\BIBdecl

\bibitem{DCC}
Y.~{Nakahara} and T.~{Matsushima}, ``A stochastic model of block segmentation
  based on the quadtree and the {Bayes} code for it,'' in \emph{2020 Data
  Compression Conference (DCC)}, 2020, pp. 293--302.

\bibitem{Shannon}
C.~E. Shannon, ``A mathematical theory of communication,'' \emph{The Bell
  System Technical Journal}, vol.~27, no.~3, pp. 379--423, 1948.

\bibitem{Huffman}
D.~A. {Huffman}, ``A method for the construction of minimum-redundancy codes,''
  \emph{Proceedings of the IRE}, vol.~40, no.~9, pp. 1098--1101, 1952.

\bibitem{arithmetic}
J.~{Rissanen} and G.~{Langdon}, ``Universal modeling and coding,'' \emph{IEEE
  Transactions on Information Theory}, vol.~27, no.~1, pp. 12--23, 1981.

\bibitem{Davisson}
L.~{Davisson}, ``Universal noiseless coding,'' \emph{IEEE Transactions on
  Information Theory}, vol.~19, no.~6, pp. 783--795, 1973.

\bibitem{enumerative}
T.~{Cover}, ``Enumerative source encoding,'' \emph{IEEE Transactions on
  Information Theory}, vol.~19, no.~1, pp. 73--77, 1973.

\bibitem{CT}
T.~{Matsushima} and S.~{Hirasawa}, ``Reducing the space complexity of a {Bayes}
  coding algorithm using an expanded context tree,'' in \emph{2009 IEEE
  International Symposium on Information Theory}, June 2009, pp. 719--723.

\bibitem{CTW}
F.~M.~J. {Willems}, Y.~M. {Shtarkov}, and T.~J. {Tjalkens}, ``The context-tree
  weighting method: basic properties,'' \emph{IEEE Transactions on Information
  Theory}, vol.~41, no.~3, pp. 653--664, 1995.

\bibitem{kontoyiannis}
\BIBentryALTinterwordspacing
I.~Kontoyiannis, L.~Mertzanis, A.~Panotopoulou, I.~Papageorgiou, and
  M.~Skoularidou, ``Bayesian context trees: Modelling and exact inference for
  discrete time series,'' arXiv, 2020. [Online]. Available:
  \url{https://arxiv.org/abs/2007.14900}
\BIBentrySTDinterwordspacing

\bibitem{JPEGLS}
M.~J. {Weinberger}, G.~{Seroussi}, and G.~{Sapiro}, ``The {LOCO-I} lossless
  image compression algorithm: principles and standardization into {JPEG-LS},''
  \emph{IEEE Transactions on Image Processing}, vol.~9, no.~8, pp. 1309--1324,
  Aug 2000.

\bibitem{kuroki}
N.~Kuroki, T.~Nomura, M.~Tomita, and K.~Hirano, ``Lossless image compression by
  two-dimensional linear prediction with variable coefficients,'' \emph{IEICE
  TRANSACTIONS on Fundamentals of Electronics, Communications and Computer
  Sciences}, vol.~75, no.~7, pp. 882--889, 1992.

\bibitem{2DAR}
X.~Wu, E.~Barthel, and W.~Zhang, ``Piecewise 2d autoregression for predictive
  image coding,'' in \emph{Proceedings 1998 International Conference on Image
  Processing. ICIP98 (Cat. No.98CB36269)}, 1998, pp. 901--904.

\bibitem{Glicbawls}
\BIBentryALTinterwordspacing
B.~Meyer and P.~Tischer, ``Glicbawls --- grey level image compression by
  adaptive weighted least squares,'' in \emph{Data Compression
  Conference}.\hskip 1em plus 0.5em minus 0.4em\relax Los Alamitos, CA, USA:
  IEEE Computer Society, mar 2001, p. 0503. [Online]. Available:
  \url{https://doi.ieeecomputersociety.org/10.1109/DCC.2001.10020}
\BIBentrySTDinterwordspacing

\bibitem{WLS}
H.~Ye, G.~Deng, J.~C. Devlin \emph{et~al.}, ``A weighted least squares method
  for adaptive prediction in lossless image compression,'' in \emph{Proc.
  Picture Coding Symp}, 2003, pp. 489--493.

\bibitem{BayesLR}
J.~Liu, G.~Zhai, X.~Yang, and L.~Chen, ``Lossless predictive coding for images
  with bayesian treatment,'' \emph{IEEE Transactions on Image Processing},
  vol.~23, no.~12, pp. 5519--5530, 2014.

\bibitem{Vanilc}
A.~Weinlich, P.~Amon, A.~Hutter, and A.~Kaup, ``Probability distribution
  estimation for autoregressive pixel-predictive image coding,'' \emph{IEEE
  Transactions on Image Processing}, vol.~25, no.~3, pp. 1382--1395, 2016.

\bibitem{TMW}
B.~Meyer and P.~Tischer, ``Tmw - a new method for lossless image compression,''
  in \emph{Proc. of 1997 Picture Coding Symposium (PCS'97)}, 1997, pp.
  533--538.

\bibitem{Bayes_avg}
A.~Martchenko and G.~Deng, ``Bayesian predictor combination for lossless image
  compression,'' \emph{IEEE Transactions on Image Processing}, vol.~22, no.~12,
  pp. 5263--5270, 2013.

\bibitem{MRP}
I.~{Matsuda}, N.~{Ozaki}, Y.~{Umezu}, and S.~{Itoh}, ``Lossless coding using
  variable block-size adaptive prediction optimized for each image,'' in
  \emph{2005 13th European Signal Processing Conference}, 2005, pp. 1--4.

\bibitem{ulacha}
\BIBentryALTinterwordspacing
G.~Ulacha, R.~Stasi?ski, and C.~Wernik, ``Extended multi wls method for
  lossless image coding,'' \emph{Entropy}, vol.~22, no.~9, 2020. [Online].
  Available: \url{https://www.mdpi.com/1099-4300/22/9/919}
\BIBentrySTDinterwordspacing

\bibitem{deep_learning}
F.~Mentzer, E.~Agustsson, M.~Tschannen, R.~Timofte, and L.~Van~Gool,
  ``Practical full resolution learned lossless image compression,'' in
  \emph{2019 IEEE/CVF Conference on Computer Vision and Pattern Recognition
  (CVPR)}, 2019, pp. 10\,621--10\,630.

\bibitem{CALIC}
X.~{Wu} and N.~{Memon}, ``Context-based, adaptive, lossless image coding,''
  \emph{IEEE Transactions on Communications}, vol.~45, no.~4, pp. 437--444,
  April 1997.

\bibitem{ISITA}
Y.~{Nakahara} and T.~{Matsushima}, ``Autoregressive image generative models
  with normal and t-distributed noise and the bayes codes for them,'' in
  \emph{2020 International Symposium on Information Theory and Its Applications
  (ISITA)}, 2020, pp. 81--85.

\bibitem{IWAIT}
\BIBentryALTinterwordspacing
Y.~Nakahara and T.~Matsushima, ``{Bayes code for two-dimensional
  auto-regressive hidden Markov model and its application to lossless image
  compression},'' in \emph{International Workshop on Advanced Imaging
  Technology (IWAIT) 2020}, P.~Y. Lau and M.~Shobri, Eds., vol. 11515,
  International Society for Optics and Photonics.\hskip 1em plus 0.5em minus
  0.4em\relax SPIE, 2020, pp. 330 -- 335. [Online]. Available:
  \url{https://doi.org/10.1117/12.2566943}
\BIBentrySTDinterwordspacing

\bibitem{H265}
G.~J. {Sullivan}, J.~{Ohm}, W.~{Han}, and T.~{Wiegand}, ``Overview of the high
  efficiency video coding ({HEVC}) standard,'' \emph{IEEE Transactions on
  Circuits and Systems for Video Technology}, vol.~22, no.~12, pp. 1649--1668,
  Dec 2012.

\bibitem{Bayes}
T.~{Matsushima}, H.~{Inazumi}, and S.~{Hirasawa}, ``A class of distortionless
  codes designed by {Bayes} decision theory,'' \emph{IEEE Transactions on
  Information Theory}, vol.~37, no.~5, pp. 1288--1293, Sep. 1991.

\bibitem{Bayes_Analysis}
B.~S. {Clarke} and A.~R. {Barron}, ``Information-theoretic asymptotics of
  {Bayes} methods,'' \emph{IEEE Transactions on Information Theory}, vol.~36,
  no.~3, pp. 453--471, May 1990.

\bibitem{range_coder}
G.~Mart{\'\i}n, ``Range encoding: an algorithm for removing redundancy from a
  digitised message,'' in \emph{Video and Data Recording Conference,
  Southampton, 1979}, 1979, pp. 24--27.

\bibitem{Waterloo}
``Image repository of the {University of Waterloo},''
  \url{http://links.uwaterloo.ca/Repository.html}, accessed: 2021-3-13.

\end{thebibliography}

\appendices

\section{validity of the proposed algorithm}\label{validity}

\subsection{The property of the model prior $p(m)$}\label{app_model_prior}

First, we prove the following lemma for a general case.

\begin{lemm}\label{subtree_decomposition}
Consider the $k$-ary complete tree $\tilde{T}$ with its depth $D$, in which each node $u$ has a parameter $g_u \in [0, 1]$. Let $\mathcal{T}$ denote the set of full subtrees which contain the root node $\lambda$ of $\tilde{T}$. Then, the following holds.
\begin{align}
\sum_{T \in \mathcal{T}} \left( \prod_{u \in \mathcal{L}^T} (1-g_u) \prod_{u' \in \mathcal{I}^T} g_{u'} \right) = 1, \label{A1}
\end{align}
where $\mathcal{L}^T$ and $\mathcal{I}^T$ denote the set of leaf nodes and inner nodes of $T$, respectively, and $g_u = 0$ for $u$ whose depth is $D$
\end{lemm}

\noindent \textit{Proof: }
Lemma \ref{subtree_decomposition} is proved by induction with respect to the depth $D$. Let $[\lambda]$ denote the tree which consists of only the root node $\lambda$ of $\tilde{T}$. When $D = 0$,
\begin{align}
\sum_{T \in \mathcal{T}} \left( \prod_{u \in \mathcal{L}^T} (1-g_u) \prod_{u' \in \mathcal{I}^T} g_{u'} \right) &= \prod_{u \in \mathcal{L}^{[\lambda]}} (1-g_u) \prod_{u' \in \mathcal{I}^{[\lambda]}} g_{u'} \label{a2}\\
&= 1-g_\lambda \label{a3}\\
&= 1, \label{a4}
\end{align}
where (\ref{a2}) is because $\mathcal{T} = \{ [\lambda] \}$, (\ref{a3}) is because $\mathcal{L}^{[\lambda]} = \{ \lambda \}$ and $\mathcal{I}^{[\lambda]} = \emptyset$, and (\ref{a4}) is because the assumption of the statement, that is $g_u = 0$ for $u$ whose depth is $D$.

If we assume (\ref{A1}) for $D = d \geq 0$ as the induction hypothesis, then the following holds for $D = d+1$.
\begin{align}
\sum_{T \in \mathcal{T}} \left( \prod_{u \in \mathcal{L}^T} (1-g_{u}) \prod_{u' \in \mathcal{I}^T} g_{u'}  \right) &= (1-g_{\lambda}) + \sum_{T \in \mathcal{T} \setminus \{ [\lambda] \}} \left( \prod_{u \in \mathcal{L}^T} (1-g_u) \prod_{u' \in \mathcal{I}^T} g_{u'} \right) \\
&= (1-g_{\lambda}) + g_\lambda  \sum_{T \in \mathcal{T} \setminus \{ [\lambda] \}}  \left( \prod_{u \in \mathcal{L}^T} (1-g_u)  \prod_{u' \in \mathcal{I}^T \setminus \{ \lambda \}}  g_{u'} \right). \label{root_decomposition}
\end{align}

Since each subtree $T \in \mathcal{T} \setminus \{ [\lambda] \}$ is identified by $k$ sub-subtrees whose root nodes are the child nodes of $\lambda$, let $\lambda_{\mathrm{child}, i}$ denote the $i$-th child node of $\lambda$ for $0 \leq i \leq k-1$ and $\mathcal{T}^{\lambda_{\mathrm{child}, i}}$ denote the set of sub-subtrees whose root node is $\lambda_{\mathrm{child}, i}$. Then, the summation in (\ref{root_decomposition}) are factorized as follows.
\begin{align}
&\sum_{T \in \mathcal{T} \setminus \{ [\lambda] \}} \left( \prod_{u \in \mathcal{L}^T} (1-g_u) \prod_{u' \in \mathcal{I}^T \setminus \{ \lambda \}} g_{u'} \right) \label{from} \\
&=\sum_{T_0 \in \mathcal{T}^{\lambda_{\mathrm{child}, 0}}} \cdots \sum_{T_{k-1} \in \mathcal{T}^{\lambda_{\mathrm{child}, k-1}}} \left\{ \left( \prod_{u \in \mathcal{L}^{T_0}} (1-g_u) \prod_{u' \in \mathcal{I}^{T_0}} g_{u'} \right) \times \cdots \times \left( \prod_{u \in \mathcal{L}^{T_{k-1}}} (1-g_u) \prod_{u' \in \mathcal{I}^{T_{k-1}}} g_{u'} \right) \right\} \\
&=\left\{ \sum_{T_0 \in \mathcal{T}^{\lambda_{\mathrm{child}, 0}}} \left( \prod_{u \in \mathcal{L}^{T_0}} (1-g_u) \prod_{u' \in \mathcal{I}^{T_0}} g_{u'} \right) \right\} \times \cdots \times \left\{ \sum_{T_{k-1} \in \mathcal{T}^{\lambda_{\mathrm{child}, k-1}}} \left( \prod_{u \in \mathcal{L}^{T_{k-1}}} (1-g_u) \prod_{u' \in \mathcal{I}^{T_{k-1}}} g_{u'} \right) \right\}. \label{to}
\end{align}

Using (\ref{A1}) for $D = d$ as the induction hypothesis,
\begin{align}
\sum_{T_{i} \in \mathcal{T}^{\lambda_{\mathrm{child}, i}}} \left( \prod_{u \in \mathcal{L}^{T_{i}}} (1-g_u) \prod_{u' \in \mathcal{I}^{T_{i}}} g_{u'} \right) = 1
\end{align}
for $0 \leq i \leq k-1$. Then,
\begin{align}
\text{(\ref{root_decomposition})} = (1-g_{\lambda}) + g_\lambda \cdot 1^k = 1.
\end{align}
Therefore, Lemma \ref{subtree_decomposition} holds for any $D$.
\hfill $\Box$

Using this lemma, the following corollaries hold for our model.
\begin{cor}\label{sum1}
The prior assumed in Assumption \ref{m_prior} satisfies $\sum_{m \in \mathcal{M}} p(m) = 1$.
\end{cor}

\begin{cor}\label{ps}
Under Assumption \ref{m_prior} and for any $s \in \mathcal{S}$, 
\begin{align}
\sum_{m \in \{ m' \in \mathcal{M} \mid s \in \mathcal{L}^{m'} \} } p(m) = (1-g_s) \prod_{s' \in \mathcal{A}_s} g_{s'}, \label{path_sum}
\end{align}
where $\mathcal{A}_s$ denotes the set of the ancestor nodes of $s$. (Let $\mathcal{A}_{s_\lambda}$ be the empty set.)
\end{cor}

\noindent \textit{Proof of Corollary 2: } Since each $m \in \{ m' \in \mathcal{M} \mid s \in \mathcal{L}^{m'} \}$ has the right hand side of (\ref{path_sum}) as the factor in its prior,
\begin{align}
\sum_{m \in \{ m' \in \mathcal{M} \mid s \in \mathcal{L}^{m'} \} } p(m) = (1-g_s) \prod_{s' \in \mathcal{A}_s} g_{s'} \sum_{m \in \{ m' \in \mathcal{M} \mid s \in \mathcal{L}^{m'} \} } \left( \prod_{s' \in \mathcal{L}^m \setminus \{ s \}} (1-g_{s'}) \prod_{s'' \in \mathcal{I}^m \setminus \{ \mathcal{A}_s \}} g_{s''} \right) .
\end{align}
Then, factorizing the sum in a similar manner from (\ref{from}) to (\ref{to}) and using Lemma \ref{subtree_decomposition} for the subtrees whose root nodes are out of $\mathcal{A}_s$, Corollary \ref{ps} is proved.

\begin{figure}[t]
\centering
\includegraphics[width=2in, clip]{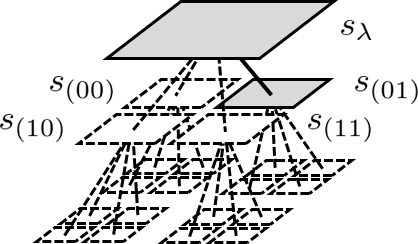}
\caption{The example for the proof of Corollary \ref{ps}} 
\label{ps_example} 
\end{figure}

For example, Fig. \ref{ps_example} shows the case where $d_\mathrm{max} = 2$, $s = s_{(01)}$, $\mathcal{A}_{s_{(01)}} = \{ s_\lambda \}$. Let $\mathcal{M}^s$ denote a set of full quadtrees whose root node is $s$. In this case, we can factorize the sum as follows.
\begin{align}
&      \sum_{m \in \{ m' \in \mathcal{M} \mid s_{(01)} \in \mathcal{L}^{m'} \} }  \left( \prod_{s \in \mathcal{L}^m \setminus \{ s_{(01)} \}}    (1-g_s)    \prod_{s' \in \mathcal{I}^m \setminus \{ \mathcal{A}_{s_{(01)}} \}}    g_{s'} \right) \nonumber \\
&= \sum_{m_{00} \in \mathcal{M}^{s_{(00)}}} \sum_{m_{10} \in \mathcal{M}^{s_{(10)}}} \sum_{m_{11} \in \mathcal{M}^{s_{(11)}}} \left\{ \left( \prod_{s \in \mathcal{L}^{m_{00}}} (1-g_{s}) \prod_{s' \in \mathcal{I}^{m_{00}}} g_{s'} \right) \right. \nonumber \\
&\left. \qquad \qquad \qquad \qquad \qquad \qquad \qquad \qquad \quad \times \left( \prod_{s \in \mathcal{L}^{m_{10}}} (1-g_{s}) \prod_{s' \in \mathcal{I}^{m_{10}}} g_{s'} \right) \left( \prod_{s \in \mathcal{L}^{m_{11}}} (1-g_{s}) \prod_{s' \in \mathcal{I}^{m_{11}}} g_{s'} \right) \right\} \\
&= \left\{ \sum_{m_{00} \in \mathcal{M}^{s_{(00)}}} \left( \prod_{s \in \mathcal{L}^{m_{00}}} (1-g_{s}) \prod_{s' \in \mathcal{I}^{m_{00}}} g_{s'} \right) \right\} \nonumber \\
&\qquad \times \left\{ \sum_{m_{10} \in \mathcal{M}^{s_{(10)}}} \left( \prod_{s \in \mathcal{L}^{m_{10}}} (1-g_{s}) \prod_{s' \in \mathcal{I}^{m_{10}}} g_{s'} \right) \right\} \left\{ \sum_{m_{11} \in \mathcal{M}^{s_{(11)}}} \left( \prod_{s \in \mathcal{L}^{m_{11}}} (1-g_{s}) \prod_{s' \in \mathcal{I}^{m_{11}}} g_{s'} \right) \right\}  \\
&= \left\{ (1-g_{s_{(00)}}) + g_{s_{(00)}}(1-g_{s_{(00)(00)}})(1-g_{s_{(00)(01)}}) (1-g_{s_{(00)(10)}})(1-g_{s_{(00)(11)}}) \right\} \nonumber \\
&\quad \times \left\{ (1-g_{s_{(10)}}) + g_{s_{(10)}}(1-g_{s_{(10)(00)}})(1-g_{s_{(10)(01)}}) (1-g_{s_{(10)(10)}})(1-g_{s_{(10)(11)}}) \right\} \nonumber \\
&\quad \times \left\{ (1-g_{s_{(11)}}) + g_{s_{(11)}}(1-g_{s_{(11)(00)}})(1-g_{s_{(11)(01)}}) (1-g_{s_{(11)(10)}})(1-g_{s_{(11)(11)}}) \right\} \\
&= 1 \cdot 1 \cdot 1 = 1.
\end{align}
\hfill $\Box$

\subsection{Proof of Lemma \ref{param_posterior}}
\begin{align}
p(v_t | v^{t-1}, m) 
&= \int p(v_t | v^{t-1}, \bm \theta^m, m) p(\bm \theta^m | v^{t-1}, m) \mathrm{d} \bm \theta^m \\
&\propto \int p(v_t | v^{t-1}, \bm \theta^m, m) p(v^{t-1} | \bm \theta^m, m) p(\bm \theta^m | m) \mathrm{d} \bm \theta^m \\
&= \int p(v_t | v^{t-1}, \theta_s^m) \int p(v^{t-1} | \bm \theta^m, m) p (\bm \theta^m | m) \mathrm{d} \bm \theta^m_{\setminus s} \mathrm{d} \theta_s^m \\
&\propto \int p(v_t | v^{t-1}, \theta_s^m) p_s (\theta_s^m) \prod_{i \in \{ i' \leq t \mid (x(i'), y(i')) \in s \}} p(v_i | v^{i-1}, \theta_s^m) \mathrm{d} \theta_s^m, \label{independent_m}
\end{align}
where $\propto$ means that the left hand side is proportional to the right hand side, regarding the variables except $v_t$ as constant, and $\bm \theta_{\setminus s}^m$ denotes the parameters $\bm \theta^m$ except $\theta_s^m$. 
Here, we used Assumptions \ref{out_prob} and \ref{param}. As a result, the formula (\ref{independent_m}) is independent of $m$. \hfill $\Box$

\subsection{Proof of Theorem \ref{alg}}
We prove the following two equations simultaneously.
\begin{align}
p(m | v^{t-1}) &= \prod_{s \in \mathcal{L}^m} (1-g_{s|t-1}) \prod_{s' \in \mathcal{I}^m} g_{s'|t-1}, \label{model_posterior} \\
q^* (v_t | v^{t-1}) &= q(v_t | v^{t-1}, s_\lambda). \label{p_c^*}
\end{align}
(\ref{model_posterior}) means that the posterior distribution of the model $m$ has the same form as the prior. (\ref{p_c^*}) is equivalent to Theorem \ref{alg}.

They are proved by induction with respect to $t$. Therefore, the proof consists of the following four steps.
\begin{description}
\item[Step 1] We prove (\ref{model_posterior}) for $t=0$.
\item[Step 2] We prove (\ref{p_c^*}) for $t=0$.
\item[Step 3] We prove (\ref{model_posterior}) for $t=k+1$ under the assumptions of (\ref{model_posterior}) and (\ref{p_c^*}) for $t=k$.
\item[Step 4] We prove (\ref{p_c^*}) for $t=k+1$ under the assumptions of (\ref{model_posterior}) for $t=k+1$ and (\ref{p_c^*}) for $t=k$.
\end{description}

\noindent \textbf{Step 1}: (\ref{model_posterior}) holds for $t=0$ because it is Assumption \ref{m_prior} itself.

\noindent \textbf{Step 2}: For $t=0$, (\ref{p_c^*}) can be proved as follows:
\begin{align}
q^* (v_0) &= \sum_{m \in \mathcal{M}} p(m)  \int  p(v_0 | \bm \theta^m, m) p(\bm \theta^m | m) \mathrm{d}\bm \theta^m \\
&= \sum_{s \in \mathcal{S}_0} \sum_{m \in \{ m' \in \mathcal{M} \mid s \in \mathcal{L}^{m'} \} }     p(m)   \int  p(v_0 | \bm \theta^m, m) p(\bm \theta^m | m) \mathrm{d}\bm \theta^m \\
&= \sum_{s \in \mathcal{S}_0} \sum_{m \in \{ m' \in \mathcal{M} \mid s \in \mathcal{L}^{m'} \} } p(m) \tilde{q}(v_0 | s)  \label{using_param_posterior} \\
&= \sum_{s \in \mathcal{S}_0} \tilde{q}(v_0 | s) \sum_{m \in \{ m' \in \mathcal{M} \mid s \in \mathcal{L}^{m'} \} } p(m) \label{decomposed_sum}\\
&=\sum_{s \in \mathcal{S}_0} \tilde{q} (v_0 | s) (1-g_s) \prod_{s' \in \mathcal{A}_s} g_{s'} \label{using_ps} \\
&= (1-g_{s_\lambda}) \tilde{q}(v_0 | s_\lambda) +   \sum_{s \in \mathcal{S}_0 \setminus \{ s_\lambda \}}   \tilde{q}(v_0 | s) (1-g_s) \prod_{s' \in \mathcal{A}_s} g_{s'} \\
&= (1-g_{s_\lambda}) \tilde{q}(v_0 | s_\lambda) + g_{s_\lambda}   \sum_{s \in \mathcal{S}_0 \setminus \{ s_\lambda \}}   \tilde{q}(v_0 | s) (1-g_s)   \prod_{s' \in \mathcal{A}_s \setminus \{ s_\lambda \}}   g_{s'}. \label{t=0}
\end{align}
Here, we used Lemma \ref{param_posterior} and Corollary \ref{ps} in (\ref{using_param_posterior}) and (\ref{using_ps}), respectively. The recursive structure in (\ref{using_ps}) and (\ref{t=0}) coincides with $q (v_0 | s_\lambda)$.


\noindent \textbf{Step 3}: In the following, we assume (\ref{model_posterior}) and (\ref{p_c^*}) for $t = k$ as the induction hypotheses. Let $r \in \mathcal{L}^m$ satisfy $(x(k), y(k)) \in r$. Then, for $t = k+1$,
\begin{align}
\prod_{s \in \mathcal{L}^m}(1-g_{s|k}) \prod_{s' \in \mathcal{I}^m} g_{s'|k}
&= \prod_{s \in \mathcal{L}^m \cap \mathcal{S}_k} (1-g_{s|k}) \prod_{s' \in \mathcal{I}^m \cap \mathcal{S}_k} g_{s'|k} \prod_{s'' \in \mathcal{L}^m \setminus \mathcal{S}_k} (1-g_{s''|k}) \prod_{s''' \in \mathcal{I}^m \setminus \mathcal{S}_k} g_{s'''|k} \\
&= (1-g_{r|k})  \prod_{s \in \mathcal{A}_r}  g_{s|k}    \prod_{s' \in \mathcal{L}^m \setminus \mathcal{S}_k}    (1-g_{s'|k})    \prod_{s'' \in \mathcal{I}^m \setminus \mathcal{S}_k}    g_{s''|k}. \label{after_update}
\end{align}

When $|r| = 1$, substituting (\ref{update}) and (\ref{recursion}) in this order,
\begin{align}
(1-g_{r|k}) \prod_{s \in \mathcal{A}_r} g_{s|k}
&= (1-g_{r|k-1}) \prod_{s \in \mathcal{A}_r} \frac{q(v_k | v^{k-1}, s_\mathrm{child})}{q (v_k | v^{k-1}, s)} g_{s | k-1}\\
&= \frac{\tilde{q}(v_{k}|v^{k-1}, r)}{q(v_{k} | v^{k-1}, s_\lambda)} (1-g_{r | k-1}) \prod_{s \in \mathcal{A}_r} g_{s | k-1}. \label{|r|=1}
\end{align}
When $|r| > 1$, substituting (\ref{update}) and (\ref{recursion}) in this order,
\begin{align}
&(1-g_{r|k}) \prod_{s \in \mathcal{A}_r} g_{s|k} \nonumber \\
&= \left( 1-\frac{q(v_k | v^{k-1}, r_\mathrm{child})}{q(v_k | v^{k-1}, r)} g_{r|k-1} \right) \prod_{s \in \mathcal{A}_r} \frac{q(v_k | v^{k-1}, s_\mathrm{child})}{q (v_k | v^{k-1}, s)} g_{s | k-1} \\
&= \left( \frac{q(v_k | v^{k-1}, r) - q(v_k | v^{k-1}, r_\mathrm{child}) g_{r|k-1}}{q(v_k | v^{k-1}, r)} \right) \prod_{s \in \mathcal{A}_r} \frac{q(v_k | v^{k-1}, s_\mathrm{child})}{q (v_k | v^{k-1}, s)} g_{s | k-1}\\
&= \left( \frac{(1-g_{r|k-1}) \tilde{q}(v_k | v^{k-1}, r) + q(v_k | v^{k-1}, r_\mathrm{child}) g_{r|k-1} - q(v_k | v^{k-1}, r_\mathrm{child}) g_{r|k-1}}{q(v_k | v^{k-1}, r)} \right) \prod_{s \in \mathcal{A}_r} \frac{q(v_k | v^{k-1}, s_\mathrm{child})}{q (v_k | v^{k-1}, s)} g_{s | k-1}\\
&= \left( \frac{(1-g_{r|k-1}) \tilde{q}(v_k | v^{k-1}, r)}{q(v_k | v^{k-1}, r)} \right) \prod_{s \in \mathcal{A}_r} \frac{q(v_k | v^{k-1}, s_\mathrm{child})}{q (v_k | v^{k-1}, s)} g_{s | k-1}\\
&= \frac{\tilde{q}(v_{k}|v^{k-1}, r)}{q(v_{k} | v^{k-1}, s_\lambda)} (1-g_{r | k-1}) \prod_{s \in \mathcal{A}_r} g_{s | k-1}. \label{|r|>1}
\end{align}
As a result, (\ref{|r|=1}) and (\ref{|r|>1}) have the same form.

On the other hand, applying the updating rule (\ref{update}),
\begin{align}
\prod_{s' \in \mathcal{L}^m \setminus \mathcal{S}_k} (1-g_{s'|k}) \prod_{s'' \in \mathcal{I}^m \setminus \mathcal{S}_k} g_{s''|k} = \prod_{s' \in \mathcal{L}^m \setminus \mathcal{S}_k} (1-g_{s'|k-1}) \prod_{s'' \in \mathcal{I}^m \setminus \mathcal{S}_k} g_{s''|k-1}.
\end{align}
Therefore,
\begin{align}
\text{(\ref{after_update})} &= \frac{\tilde{q}(v_{k}|v^{k-1}, r)}{q(v_{k} | v^{k-1}, s_\lambda)} (1-g_{r | k-1}) \prod_{s \in \mathcal{A}_r} g_{s | k-1} \prod_{s' \in \mathcal{L}^m \setminus \mathcal{S}_k} (1-g_{s'|k-1}) \prod_{s'' \in \mathcal{I}^m \setminus \mathcal{S}_k} g_{s''|k-1}  \\
&= \frac{\tilde{q}(v_{k}|v^{k-1}, r)}{q(v_{k} | v^{k-1}, s_\lambda)} \prod_{s \in \mathcal{L}^m} (1-g_{s|k-1}) \prod_{s' \in \mathcal{I}^m} g_{s'|k-1} \\
&= \frac{\tilde{q}(v_{k}|v^{k-1}, r)}{q^*(v_{k} | v^{k-1})} p(m | v^{k-1}) \label{using_model_posterior} \\
&= \frac{p(v_{k}|v^{k-1}, m)}{p(v_{k} | v^{k-1})} p(m | v^{k-1}) \label{using_p_c^*} \\
&= p(m | v^k).
\end{align}
In (\ref{using_model_posterior}), we used (\ref{model_posterior}) and (\ref{p_c^*}) as the induction hypothesis. In (\ref{using_p_c^*}) we used Lemma \ref{param_posterior} and Proposition \ref{ideal}. Thus, (\ref{model_posterior}) holds for $t=k+1$.

In addition, it holds that
\begin{align}
    \sum_{m \in \{ m' \in \mathcal{M} | s \in \mathcal{L}^{m'} \} }     p(m | v^k) = (1-g_{s|k}) \prod_{s' \in \mathcal{A}_s} g_{s'|k}, \label{model_posterior_sum}
\end{align}
since the posterior $p(m | v^k)$ has the same form as the prior $p(m)$ and can be applied Corollary \ref{ps}.

\noindent \textbf{Step 4}: (\ref{p_c^*}) can be proved for $t=k+1$ in a similar manner to the case where $t=0$.
%
\begin{align}
q^* (v_{k+1} | v^k) 
&=   \sum_{s \in \mathcal{S}_{k+1}}   \tilde{q}(v_{k+1} | v^k, s)     \sum_{m \in \{ m' \in \mathcal{M} | s \in \mathcal{L}^{m'} \} }     p(m | v^k) \\
&= \sum_{s \in \mathcal{S}_{k+1}}  \tilde{q} (v_t | v^k, s) (1-g_{s|k}) \prod_{s' \in \mathcal{A}_s} g_{s'|k} \label{using_ps2}\\
&= (1-g_{s_\lambda | k}) \tilde{q}(v_{k+1} | v^k, s_\lambda) + g_{s_\lambda | k}   \sum_{s \in \mathcal{S}_{k+1} \setminus \{ s_\lambda \}}   \tilde{q}(v_{k+1} | v^k, s) (1-g_{s | k})   \prod_{s' \in \mathcal{A}_s \setminus \{ s_\lambda \}}   g_{s' | k}. \label{t=k+1}
\end{align}
In (\ref{using_ps2}), we used (\ref{model_posterior_sum}). The recursive structure in (\ref{using_ps2}) and (\ref{t=k+1}) coincides with $q (v_{k+1} | v^k, s_\lambda)$.


\hfill $\Box$

\section{The algorithm to calculate $m^\mathrm{MAP}$}\label{max_algorithm}
In this appendix we derive the algorithm to calculate $\mathrm{arg} \max_{m} p(m | v^t)$. At first, $\max_{m} p(m | v^t)$ can be decomposed in a similar manner to the proof of Lemma \ref{subtree_decomposition} by replacing the sum for the max.

\begin{align}
\max_{m \in \mathcal{M}} p(m | v^t) = \max \Biggl\{ 1-g_{s_\lambda | t}, \quad & g_{s_\lambda | t} \max_{m_{00} \in \mathcal{M}^{s_{(00)}}} \Biggl\{  \prod_{s \in \mathcal{L}^{m_{00}}} (1-g_{s | t})   \prod_{s' \in \mathcal{I}^{m_{00}} \setminus \{ s_\lambda \}}   g_{s' | t} \Biggr\} \nonumber \\
&\qquad \times \max_{m_{01} \in \mathcal{M}^{s_{(01)}}} \Biggl\{  \prod_{s \in \mathcal{L}^{m_{01}}} (1-g_{s | t})   \prod_{s' \in \mathcal{I}^{m_{01}} \setminus \{ s_\lambda \}}   g_{s' | t} \Biggr\} \nonumber \\
&\qquad \times \max_{m_{10} \in \mathcal{M}^{s_{(10)}}} \Biggl\{  \prod_{s \in \mathcal{L}^{m_{10}}} (1-g_{s | t})   \prod_{s' \in \mathcal{I}^{m_{10}} \setminus \{ s_\lambda \}}   g_{s' | t} \Biggr\} \nonumber \\
&\qquad \times \max_{m_{11} \in \mathcal{M}^{s_{(11)}}} \Biggl\{  \prod_{s \in \mathcal{L}^{m_{11}}} (1-g_{s | t})   \prod_{s' \in \mathcal{I}^{m_{11}} \setminus \{ s_\lambda \}}   g_{s' | t} \Biggr\} \Biggr\} .
\end{align}

We define a recursive function $\phi_t : \mathcal{S} \to \mathbb{R}$ as follows.
\begin{defi}
\begin{align}
\phi_t (s) \coloneqq 
\begin{cases}
1, & |s| = 1\\
\max \big\{ 1-g_{s|t}, \, g_{s|t} \phi_t (s_{\mathrm{child}_{00}}) \phi_t (s_{\mathrm{child}_{01}}) \phi_t (s_{\mathrm{child}_{10}}) \phi_t (s_{\mathrm{child}_{11}}) \big\}, &\mathrm{otherwise}.
\end{cases}
\end{align}
Here, $s_{\mathrm{child}_{00}}$, $s_{\mathrm{child}_{01}}$, $s_{\mathrm{child}_{10}}$, and $s_{\mathrm{child}_{11}}$ are child nodes of $s$ of the complete quadtree on $\mathcal{S}$
\end{defi}
Then, $\max_m p(m | v^{t})$ can be calculated by $\phi_t (s_\lambda).$

Next, we define the following flag variable $h_{s|t} \in \{ 0, 1\}$.
\begin{defi}
\begin{align}
h_{s | t} \coloneqq
\begin{cases}
0, & 1-g_{s|t} \geq g_{s|t} \phi_t (s_{\mathrm{child}_{00}}) \phi_t (s_{\mathrm{child}_{01}}) \phi_t (s_{\mathrm{child}_{10}}) \phi_t (s_{\mathrm{child}_{11}})\\
1, & \mathrm{otherwise}.
\end{cases}
\end{align}
\end{defi}
We can calculate $h_{s|t}$ and $\phi_t (s)$ simultaneously. Then, $\mathrm{arg} \max_m p(m | v^t)$ is identified as the model which satisfies
\begin{align}
&s \in \mathcal{I}^m \Rightarrow h_{s|t} = 1,\\
&s \in \mathcal{L}^m \Rightarrow h_{s|t} = 0.
\end{align}
Such a model can be searched by backtracking from $s_\lambda$ after the calculation of $\phi_t (s_\lambda)$ and $h_{s_\lambda | t}$.

\ifCLASSOPTIONcaptionsoff
  \newpage
\fi

\begin{IEEEbiographynophoto}{Yuta Nakahara}
He received his B.E.\ degree in the Department of Applied Mathematics from Waseda University, Tokyo, Japan, in 2014. He received M.E.\ and Dr.E.\ degrees in the Department of Pure and Applied Mathematics from Waseda University, Tokyo, Japan, in 2016 and 2019, respectively. He has been a lecturer of Center for Data Science at Waseda University, Tokyo, Japan, since 2019.
His research interests include information theory and its applications, especially error correcting codes and lossless image compression.
\end{IEEEbiographynophoto}

\begin{IEEEbiographynophoto}{Toshiyasu Matsushima}
He received the B.E.\ degree, M.E.\ degree and Dr.E degree in Industrial Engineering and Management from Waseda University, Tokyo, Japan, in 1978, 1980 and 1991, respectively. From 1980 to 1986, he joined NEC Corporation, Kanagawa, Japan. From 1989 to 1993, he was a lecturer at the Department of Management Information, Yokohama College of Commerce. From 1993, he was an associate professor, and from 1996 to 2007 a professor at Department of Industrial and Management System Engineering, Waseda University. Since 2007, he has been a professor at Department of Applied Mathematics, Waseda University. From 2001 to 2002, he was a visiting researcher at Department of Electrical Engineering, University of Hawaii, USA. From 2011 to 2012, He was a visiting scalar at Department of Statistics, the University of California, USA. His research interests are information theory, statistics, learning theory and their applications. He is a member of the Japan Society for Quality Control, the Japan Industrial Management Association, the Japan Society for Artificial Intelligence, and IEEE.
\end{IEEEbiographynophoto}







\end{document}